\documentclass[sigconf,nonacm]{acmart}
\usepackage{enumitem}
\usepackage[tikz]{bclogo}
\usepackage{tikz}
\usetikzlibrary{calc}
\usepackage{pgfplots}
\pgfplotsset{compat=1.17}
\usetikzlibrary{positioning}
\usepackage{amsmath}
\usepackage{booktabs}
\usepackage{listings}
\usepackage{xcolor}
\usepackage{tikzscale}
\usepackage{hyperref}
\usepackage{graphicx}
\usepackage{float}
\usepackage{subcaption}
\usepackage{multirow}
\usepackage[many]{tcolorbox}
\DeclareMathOperator*{\argmax}{argmax}
\definecolor{edgeblue}{RGB}{0, 0, 200}
\definecolor{blue}{RGB}{0, 0, 255}

\theoremstyle{definition}

\newcommand{\myInsight}[2]{
    \begin{bclogo}[logo=\bclampe, couleur=blue!26, 
                arrondi = 0.1, ombre=False, barre=none]{#1 $\Rightarrow$ \normalfont{#2}}
    \end{bclogo}
}

\acmConference[ICSE 2022]{The 44th International Conference on Software Engineering}{May 21–29, 2022}{Pittsburgh, PA, USA}

\title{DeepAnalyze: Learning to Localize Crashes at Scale}
\author{Manish Shetty}
\email{t-mamola@microsoft.com}
\affiliation{%
  \institution{Microsoft Research}
  \streetaddress{}
  \city{Bangalore}
  \country{India}
}

\author{Chetan Bansal}
\email{chetanb@microsoft.com}
\affiliation{%
  \institution{Microsoft Research}
  \streetaddress{}
  \city{Redmond}
  \country{USA}
}

\author{Suman Nath}
\email{sumann@microsoft.com}
\affiliation{%
  \institution{Microsoft Research}
  \streetaddress{}
  \city{Redmond}
  \country{USA}
}

\author{Sean Bowles}
\email{sbowl@microsoft.com}
\affiliation{%
  \institution{Microsoft}
  \streetaddress{}
  \city{Redmond}
  \country{USA}
}

\author{Henry Wang}
\email{hewang@microsoft.com}
\affiliation{%
  \institution{Microsoft}
  \streetaddress{}
  \city{Redmond}
  \country{USA}
}

\author{Ozgur Arman}
\email{oarman@microsoft.com}
\affiliation{%
  \institution{Microsoft}
  \streetaddress{}
  \city{Redmond}
  \country{USA}
}

\author{Siamak Ahari}
\email{sahari@microsoft.com}
\affiliation{%
  \institution{Microsoft}
  \streetaddress{}
  \city{Redmond}
  \country{USA}
}

\begin{document}

\newcommand{\deepanalyze}{DeepAnalyze}
\newcommand{\banganalyze}{{\tt !analyze}}
\newcommand{\CompanyX}{CompanyX}
\newcommand{\Paragraph}[1]{\smallskip\noindent{\bf #1}}
\newcommand{\application}{\textbf{\textcolor{red}{application}}}

\newcommand{\todo}[1]{\textbf{\textcolor{red}{TODO: #1}}}
\newcommand{\add}[1]{\textbf{\textcolor{blue}{X #1 X}}}
\newcommand{\addRef}{\todo{REF HERE}}
\newcommand{\reduceVSpace}{\vspace{-3mm}}

\newcommand{\suman}[1]{\textcolor{blue}{\noindent{{\bf \fbox{Suman:} {\textbf{\textit{#1}}}}}}}
\newcommand{\manish}[1]{\textcolor{red}{\noindent{{\bf \fbox{Manish:} {\textbf{\textit{#1}}}}}}}

\begin{abstract}
%Crash localization at scale is a challenging problem due to vast amount of binaries, exception types, and platforms. Existing crash reporting systems such as Windows Error Reporting (WER) have been developed over decades and comprise of 100K+ LOC and a large number of rules and heuristics. Further, binary specific plugins need to be written in order to adapt the default rules. In this work, we do the first large-scale empirical study of crashes in the wild. We analyze different aspects such as the crash causing binaries, top exception classes, and also, the crash stack complexity. We then propose a novel Multi-Task sequence labeling approach for crash localization. We prove the efficacy of this approach by evaluating on real-world crashes from four popular Windows applications. Furthermore, to solve challenges with new and evolving applications with minimal crash data, we propose and evaluate a cross-application transfer learning approach for crash localization.
Crash localization, an important step in debugging crashes, is challenging when dealing with an extremely large number of diverse applications and platforms and underlying root causes. Large-scale error reporting systems, e.g., Windows Error Reporting (WER), commonly rely on manually developed rules and heuristics to localize {\em blamed frames} causing the crashes. As new applications and features are routinely introduced and existing applications are run under new environments, developing new rules and maintaining existing ones become extremely challenging.

We propose a data-driven solution to address the problem. We start with the first large-scale empirical study of $362K$ crashes and their blamed methods reported to WER by tens of thousands of applications running in the field. The analysis provides valuable insights on where and how the crashes happen and what methods to blame for the crashes. These insights enable us to develop \deepanalyze{}, a novel multi-task sequence labeling approach for identifying blamed frames in stack traces. We evaluate our model with over a million real-world crashes from four popular Microsoft applications and show that \deepanalyze{}, trained with crashes from one set of applications, not only accurately localizes crashes of the same applications, but also bootstrap crash localization for other applications with zero to very little additional training data. 

\end{abstract}


\maketitle

\begin{figure*}
    \centering
	\subfloat[\centering Crash stack 1]{
		{
			\begin{tikzpicture}[
    node distance = 0pt,
    lineno/.style = {draw=white!40, solid, 
                     minimum height=1.6em, minimum width=2em,
                     outer sep=0pt, align = left},
    frame/.style = {draw=black!40, fill=gray!10, solid, 
                     minimum height=1.5em, 
                     minimum width=33.2em,
                     text width=33.2em,
                     outer sep=0pt, align = left},  
    redframe/.style = {draw=black!40, fill=red!30, solid, 
                     minimum height=1.5em, 
                     minimum width=33.2em,
                     text width=33.2em,
                     outer sep=0pt, align = left},
        ]
    \ttfamily
    \footnotesize
    %Nodes
    \node[lineno] (l1) {0};
    \node[lineno, below=of l1] (l2) {1};
    \node[lineno, below=of l2] (l3) {2};
    \node[lineno, below=of l3] (l4) {3};
    \node[lineno, below=of l4] (l5) {4};
    \node[lineno, below=of l5] (l6) {5};
    \node[lineno, below=of l6] (l7) {6};
    \node[lineno, below=of l7] (l8) {7};

    \node[frame, right=of l1] (f0) {msedge\_elf.dll!crash\_reporter::DumpWithoutCrashing};
    
    \node[frame, right=of l2] (f1) {msedge.dll!base::debug::DumpWithoutCrashing};
    
    \node[redframe, right=of l3] (f2) {msedge.dll!gl::DirectCompositionChildSurfaceWin::ReleaseDrawTexture};
    
    \node[frame, right=of l4] (f3) {msedge.dll!gl::DirectCompositionChildSurfaceWin::SwapBuffers};
    
    \node[frame, right=of l5] (f4) {msedge.dll!gl::DirectCompositionChildSurfaceWin::SwapBuffers};
    
    \node[frame, right=of l6] (f5)
    {msedge.dll!gl::GLSurfaceAdapter::PostSubBuffer};
    
    \node[frame, right=of l7] (f6) {msedge.dll!gpu::PassThroughImageTransportSurface::PostSubBuffer};
    
    \node[frame, right=of l8] (f7) {\ldots};
\end{tikzpicture}}
		    \label{fig:example-stack-1}
	}
	\subfloat[\centering Crash stack 2]{
		{
			\begin{tikzpicture}[
    node distance = 0pt,
    lineno/.style = {draw=white!40, solid, 
                     minimum height=1.6em, minimum width=2em,
                     outer sep=0pt, align = left},
    frame/.style = {draw=black!40, fill=gray!10, solid, 
                     minimum height=1.5em, 
                     minimum width=33.2em,
                     text width=33.2em,
                     outer sep=0pt, align = left},  
    redframe/.style = {draw=black!40, fill=red!30, solid, 
                     minimum height=1.5em, 
                     minimum width=33.2em,
                     text width=33.2em,
                     outer sep=0pt, align = left},
    yellowframe/.style = {draw=black!40, fill=yellow!30, solid, 
                 minimum height=1.5em, 
                 minimum width=33.2em,
                 text width=33.2em,
                 outer sep=0pt, align = left},
        ]
    \ttfamily
    \footnotesize
    %Nodes
    \node[lineno] (l1) {0};
    \node[lineno, below=of l1] (l2) {1};
    \node[lineno, below=of l2] (l3) {2};
    \node[lineno, below=of l3] (l4) {3};
    \node[lineno, below=of l4] (l5) {4};
    \node[lineno, below=of l5] (l6) {5};
    \node[lineno, below=of l6] (l7) {6};
    \node[lineno, below=of l7] (l8) {7};
    
    \node[redframe, right=of l1] (f0) {igd10iumd64.dll!OpenAdapter10\_2};
    
    \node[frame, right=of l2] (f1) {d3d11.dll!NDXGI::CDevice::RotateResourceIdentities};
    
    \node[frame, right=of l3] (f2) {dxgi.dll!CDXGISwapChain::PresentImplCore};
    
    \node[frame, right=of l4] (f3) {dxgi.dll!CDXGISwapChain::PresentImpl};
    
    \node[frame, right=of l5] (f4) {dxgi.dll!CDXGISwapChain::[IDXGISwapChain4]::Present1};
    
    \node[yellowframe, right=of l6] (f5) {msedge.dll!gl::DirectCompositionChildSurfaceWin::ReleaseDrawTexture};
    
    \node[frame, right=of l7] (f6) {msedge.dll!gl::DirectCompositionChildSurfaceWin::SwapBuffers};
    
    \node[frame, right=of l8] (f7) {\ldots};
\end{tikzpicture}}
		    \label{fig:example-stack-2}
	}
	\vspace{-2mm}
	\caption{Examples of crash stacks from Microsoft Edge and their crash 
	locations (red frame)}%
	\label{fig:example-stack}
	\vspace{-4mm}
\end{figure*}

\section{Introduction}
When software crashes in the wild, often the primary sources of information available for debugging are {\em crash stacks} -- stack traces collected during the crashes~\cite{schroter2010stack}. A crash stack contains what methods were executing during a crash. It may also contain other valuable information such as executed binaries and code locations that can hint as to what might have caused the crash. Due to its importance, many error reporting systems, e.g., Windows Error Reporting (WER)~\cite{glerum2009debugging}, Apple Crash Reporter~\cite{applecrash}, Mozilla Crash Reporter~\cite{mozillacrash}, Chrome Crash Reporter~\cite{chromiumcrash}, have been deployed to automate the collection of crash stacks (along with other information, such as memory dumps). 
An important step in investigating a crash is {\em crash localization}: 
identifying the method in the crash stack that contains, or is the closest to,\footnote{When a stack trace does not contain the true crash location~\cite{gu2019does}, we consider identifying the method closest to the crash location in 
caller-callee relationship.} the crash location. We denote such a method as 
the {\em blamed method}, and the stack frame containing it as the {\em blamed frame}. Blamed methods play an important role in organizing crash reports into ``buckets'' (i.e., categories), which in turn help developers prioritize frequently seen buckets~\cite{glerum2009debugging}. Moreover, investigation of a crash often starts from the blamed method as it helps developers isolate the crash location.

Crash localization needs to be automated in large-scale error reporting systems, such as WER, as they receive millions of crash reports per day \cite{glerum2009debugging}. A common practice \cite{glerum2009debugging,ubuntucrash} is to use a collection of manually written heuristics such as ``Never mark 
\texttt{Foo()} as a blamed method'', ``\texttt{Bar()} can be a blamed method only if the symbol \texttt{Baz} appears in the crash stack'', and so on. The system then scans through the frames of a crash stack to identify a blamed method that is consistent with these rules. Ideally, the rules should be consistent (i.e., not contradictory) with each other, and have good accuracy and coverage. This is nontrivial for large and evolving systems; when a new feature or application is introduced, someone with good domain knowledge needs to write new application-specific rules. WER currently has 50+ such application-specific libraries of rules. 

Prior work in crash analysis has heavily focused on crash bucketization \cite{dang2012rebucket, glerum2009debugging, van2018semantic}. But, in order to do effective bucketization, crash localization is critical. Wu et al. \cite{wu2014crashlocator} proposed CrashLocator which leverages source code along with the static call graph for crash localization. However, this is not feasible at the scale of WER that needs to localize crashes for a multitude of applications in the wild. This paper addresses these limitations with a fresh data-driven approach. 
Inspired by the abundance of data in existing error reporting 
systems and recent advancement in deep learning techniques, we
demonstrate how to learn from past crashes to identify the
blamed method in a new crash stack with high accuracy.

As a first step towards our data-driven approach, we analyze $\approx362K$ crash stacks collected by WER from $\approx8.7K$ software components. Our analysis highlights the huge diversity of crash stacks: they come from many different first and third party binaries, and from many different methods and namespaces within each binary. The underlying \textit{problem classes}, which denote high level root cause types such as heap corruption or stack overflow, are also diverse. Finally, a crash trace often contains many methods, only one of which needs to be identified as the blamed method. All of the above highlight the challenges of manually developing and maintaining application-specific crash-localization heuristics.

Our analysis also provides several insights that guide our choice of an 
effective machine-learning solution. We first tried a linear binary classifier (Logistic Regression) that, given features of an individual 
frame, predicts the likelihood of it being the blamed frame. As we show in 
Section \ref{sec:rq1-eval}, this simple model, however, did not work well for many problem classes. Upon further analysis of our dataset, we found that {\em the context in which a method appears in a crash stack (e.g., methods that appear before and after it) plays an important role in it being a blamed
method or not}. Consequently, a method can be the blamed method in one crash 
stack but not in another. This is illustrated with two 
crash traces from Microsoft Edge, shown in Figure \ref{fig:example-stack}. 
Both the traces contain Edge's method \texttt{ReleaseDrawTexture}, but WER 
identifies it as the blamed method only for the first trace. In the first trace, methods around \texttt{ReleaseDrawTexture} are relatively less crash-prone based on their past history (e.g., the logging methods above it). In the second trace, however, \texttt{ReleaseDrawTexture} has more crash-prone methods from a user mode Intel graphics driver, one of which is blamed by WER. 

We use this insight on the importance of context in a novel formulation of crash localization as {\em sequence labeling}. Sequence 
labeling, widely used in natural language processing, uses context to assign a categorical 
label (e.g., parts of speech) to each member of a sequence. In our formulation, we treat a crash stack as a sequence of frames and aim to assign to each frame a binary category indicating whether it is a blamed frame or not. But, applying sequence labeling to crash localization requires addressing several challenges. 

First, like many other machine learning tasks, we need to select a good set of features and a suitable model that can capture context. Here, we revisit our data analysis to seek insights. Our analysis shows that even though a stack trace may contain many 
methods, only a small number of them are likely to be a crash location, e.g., methods that appear towards the top of the stack, have global semantic importance, and that are implemented in application code rather than the underlying system. We therefore extract features that summarize these properties of a stack frame.
We also observe that context for a frame depends on the call chain through which it is invoked. We capture this sequential context flow in the stack using a Bi-directional Long Short-term Memory (Bi-LSTM) layer \cite{hochreiter1997long} that interprets the stack both top-down and bottom-up.

Second, traditional sequence labeling may label multiple tokens with the same category. In our context, however, only one frame in a stack trace can be blamed. To address this, we first use an attention mechanism \cite{bahdanau2014neural} to identify sections of a stack trace that are more likely to contain the crash location. Then, we model the frame-level labeling task jointly using linear chain Conditional Random Fields (CRF) \cite{lafferty2001conditional}. 

Finally, to tackle constantly evolving software, it is important that models learned from crashes in one set of applications are useful not only for those, but also for other applications, e.g., newly released ones that have very little training data. We propose a transfer learning approach to achieve this goal. 

\begin{figure}[!ht]
	\includegraphics[scale=0.6]{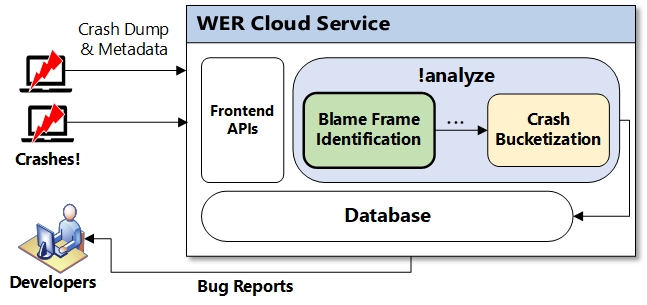}
	\vspace{-2mm}
	\caption{An overview of WER}
	\vspace{-4mm}
	\label{fig:WER}
\end{figure}

In this work, we present and package our models in a system called \textbf{\deepanalyze{}} -- a deep learning based solution for large-scale crash localization. We have evaluated \deepanalyze{} with over a million real-world crash stacks from four popular Microsoft applications (Edge, Word, Excel, and Outlook). Our results show that \deepanalyze{}'s novel multi-task sequence labeling approach has an average accuracy of 0.9 and it outperforms several baselines. Also, we show that using transfer learning, our model, learned from one set of applications, can also be effectively ($\approx 0.8$ accuracy) used for completely new and unseen applications with no new training data. Lastly, we show that these models can be easily fine-tuned to new applications, where the accuracy quickly approaches $\approx0.9$ with as little as a few thousands additional training samples.

In summary, we make the following contributions: 

\begin{enumerate}
    \item We conduct the first large-scale empirical study of crashes in the wild, and discuss new insights about crash sources, problem types and characteristics of their blamed methods.
    \item We propose \deepanalyze{}, a novel Multi-task learning based approach for crash localization only using stack traces. 
    \item We evaluate \deepanalyze{} on $4$ popular Windows applications showing that it has an average accuracy of 0.9 and outperforms several baselines.
    \item We leverage transfer learning for cross-application crash localization and demonstrate that with a small amount of data, we can localize crashes for new/unseen applications with high accuracy.
\end{enumerate}
\begin{figure*}[!ht]
	\centering
	\subfloat[\centering Top-25 crashing binaries]{
		{
			\includegraphics[scale=0.6]{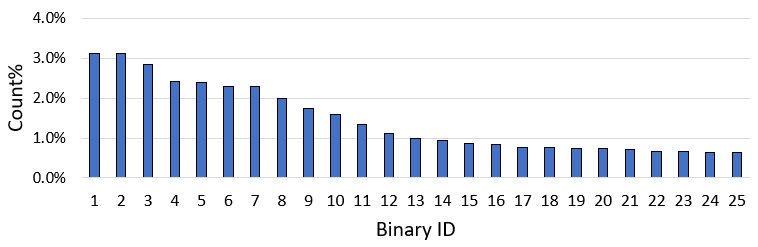}
	}}
	\hspace{4em}
	\subfloat[\centering Types of Top-100 crashing software components]{
		{
			\includegraphics[scale=0.8]{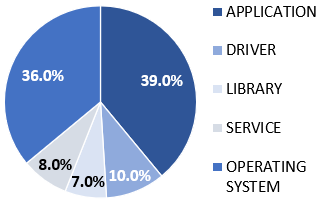}
	}}
	\caption{Distribution of crashing binaries and software components}
	\label{fig:crashingbinaries}
	\vspace{-3mm}
\end{figure*}

\section{Background}
\label{sec:background}
\subsection{Windows Error Reporting}
%Due to the large ecosystem of applications and the continuous updates, it's non-trivial for buggy software to be released to end users. Hence, it's important to detect and triage such issues at scale. Microsoft has built and operated WER \cite{glerum2009debugging, url-about-wer}, a distributed system for collecting bug reports for over two decades now. So far, it has collected tens of billions of bug reports from billions of devices and millions of applications. WER collects crash dumps for problems detected by the Windows OS for both user mode and kernel mode crashes. Developers can use the WER infrastructure to subscribe to crash reports for their applications. A key functionality of WER is to assign \textit{buckets} to the crashes so that similar bugs can de-duplicated and triaged together. A bucket is basically a signature to identify a unique bug. \todo{add example of bucket}. Once the number of crashes has exceeded a certain threshold, a bug report is created and it is triaged to the appropriate developer using the bucket metadata.

Released software often contains bugs that cause the software to crash in the field. To automate collecting crash information, large software companies deploy error reporting systems. For example, Microsoft has built a distributed error-reporting system called Windows Error Reporting (WER) \cite{glerum2009debugging, url-about-wer}, which has been in operation for over two decades now. When a Microsoft software (such as Windows, Word, and Edge) crashes in the field, with user permission, it sends to WER a {\em crash report}. A crash report contains crash stacks, information of the crashed application and its runtime, and optionally a memory dump collected during the crash (Figure~\ref{fig:WER}). So far, WER has collected tens of billions of crash reports from billions of devices. 

A key functionality of WER is to assign \textit{buckets} to the crashes so that similar bugs can de-duplicated and triaged together. Once the number of crashes in a bucket has exceeded a certain threshold, a bug report is created and it is triaged to the appropriate developer using the bucket metadata. A bucket is basically a signature to identify a unique bug. Here is an example of a bucket: \textit{MEMORY\_CORRUPTION\_c0000005\_contoso.exe!WriteToChild}. It contains the problem class, exception code and the blamed frame which caused the crash. The bucket is computed by analyzing a crash report, as described next.

\subsection{Crash localization in WER}
In order to analyze a crash report, WER uses \banganalyze{} \cite{url-using-analyze}, a debugger extension which uses the call stack from the crash dump along with metadata such as loaded modules, memory dump and exception code to identify blamed frame and the underlying problem class that caused the crash (e.g., out of memory, stack overflow). \banganalyze{} has been built and maintained for more than two decades, using over 200,000 lines of code and hundreds of heuristics written by domain experts. \banganalyze{} also provides an extensibility mechanism that both Microsoft and external developers use to build plug-ins for extending or overriding the default logic with application-specific rules. \banganalyze{} currently has 50+ such extensions. The extensions can be nontrivial. For example, the extension for Edge consists of over 2K LOC!

%In order to assign bucket to a crash, it first tries to identify the \textit{blame frame} in the call stack. The blame frame is identified using several heuristics and rules. 

\banganalyze{} source code has been through many years of improvement and updates for analyzing different error codes and buckets. To understand the pain points of \banganalyze{}, we talked to several application owners. At present, the rules for crash report analysis are required to be written into the code. This results in huge deployment overheads; for instance, updating or deploying new rules can range anywhere between one to three months. Further, the rules often tend to be very specific, and need to be updated as the application code base evolves. For new applications, the application owner and \banganalyze{} developers need to work together to implement the logic of the new rules into the code base. Also, bringing up a new application is usually time consuming and requires deep domain knowledge of \banganalyze{} code base. With \deepanalyze{}, our goal is to eliminate or significantly simplify this laborious and time consuming process with an agile and fast data-driven solution.

%A data-driven approach that can use historical data to to learn how to identify blame frames for a new application can significantly reduce or even eliminate this cost.
\begin{table}[!ht]
	\caption{Basic statistics of the dataset used in our study}
	\vspace{-2mm}
	\label{tab:data}
	\begin{tabular}{|l|l|}
		\hline
		\# of crash stacks & $\approx 362K$ \\	\hline
		\# of unique software components & $\approx 8.7K$ \\ \hline
		\# of unique binaries & $\approx 16.3K$ \\ \hline
		\# of unique namespaces & $\approx 38K$ \\ \hline
		\# of unique methods & $\approx 85K$ \\ \hline
		%\# of unique blame binaries & $\approx 13.2K$\\ \hline
		\# of unique blamed methods & $\approx 18K$ \\ \hline
		
	\end{tabular}
	\vspace{-4mm}
\end{table}

\begin{figure*}[!ht]
\centering
\subcaptionbox{Stack depth\label{fig:stackDepth}}{\includegraphics[height=0.2\textwidth, width=0.25\textwidth]{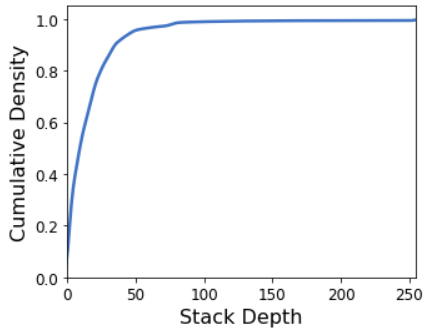}}%
\hfill % <-- Seperation
\subcaptionbox{Distinct binaries per stack\label{fig:stackBinaries}}{\includegraphics[height=0.2\textwidth, width=0.25\textwidth]{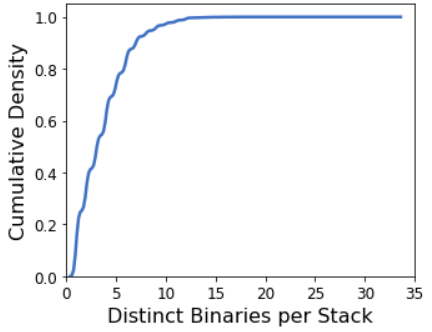}}%
\hfill % <-- Seperation
\subcaptionbox{Avg stack depth for top-25 software components \label{fig:appStackDepth}}{\includegraphics[height=0.2\textwidth,width=0.45\textwidth]{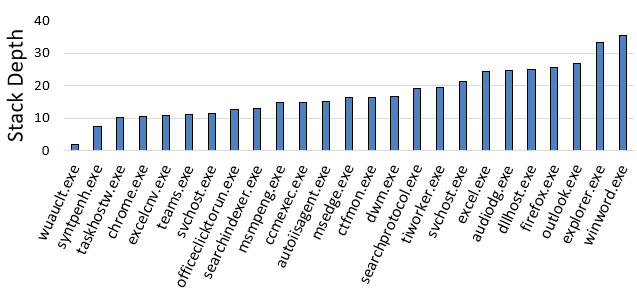}}%
\vspace{-0.5em}
\caption{Properties of crash stacks}
\vspace{-3mm}
\label{fig:crashStackProperties}
\end{figure*}

\section{Empirical Analysis of Crashes}
\label{sec:empirical-analysis}
% \suman{TODO: consistent naming. We use the term 'application' to denote drivers, systems, application, etc. Is there a better term?}
In this section, we analyze a large collection of crashes collected by WER 
to understand various properties of crash stacks. Our dataset contains  
$362,249$ unique crash stacks, uniformly sampled from crashes collected by WER in a 1 week period. We also study properties of their problem classes and blamed 
frames/methods, as determined by \banganalyze{} with its manually written heuristics. 

\subsection{Crash sources}
WER collects crash stacks from thousands of applications developed by both Microsoft and non-Microsoft developers. Our sample dataset contains crash stacks from $\approx 16.3K$ 
binaries of $\approx 8.7K$ software components\footnote{We use the term {\em 
software component} to denote various types of software systems including user-mode 
applications, OS or infrastructure systems, libraries, drivers, etc.}, including popular Microsoft applications such as Excel, Word, Outlook, Edge, etc. The crash stacks include $\approx 85K$ unique methods from $\approx 38K$ unique namespaces. Table~\ref{tab:data} summaries the statistics.

Figure~\ref{fig:crashingbinaries} shows distributions of top binaries and software components where crashes occur. The distribution has a long tail, with most applications contributing a small fraction of crash stacks. 
As shown in Figure~\ref{fig:crashingbinaries} (a), there are only 10 binaries in our dataset, each of which accounts for $>1.5\%$ (and the top binary accounts for $\approx3\%$) of the total crashes. We also analyzed the types of top 100 software components in terms of their crash frequencies. As shown in Figure~\ref{fig:crashingbinaries} (b), $75\%$ of them are user-mode applications (e.g., Microsoft Excel) or underlying 
systems (e.g., Windows Desktop Window Manager (dwm)), with the remaining $25\%$ being drivers, libraries, and services. This shows that in a large-scale error reporting system like WER, crash stacks come from many different sources. Hence, application-specific mechanisms to identify blame frames do not scale well. 
%Most rules in WER are application-independent; however, developers often encounter crashes whose blame frame cannot be accurately identified with the generic rules. This results in developers writing new ``plug-ins'' that use application-specific rules. WER currently has 50+ such plug-ins. The process of writing and maintaining plug-ins is time consuming and expensive. For instance, the Edge plug-in for \banganalyze{} is over 2K LOC. A data-driven approach that can use historical data to to learn how to identify blame frames for a new application can significantly reduce or even eliminate this cost.\\ 
\\
\myInsight{Finding \#1}
{Crash stacks come from many different sources and hence application-specific crash localization does not scale well.}

\subsection{Crash stacks}
\Paragraph{Stack depth.} Stack depth (i.e., number of frames in the stack) at the time of a crash indicates how deep the crash happens, in terms of nested method calls. Figure~\ref{fig:crashStackProperties} (a) shows the distribution of the depths (i.e., number 
of frames) of all crash stacks in our dataset. Mean and median depths are 16 and 9 respectively. While majority have fewer than 10 frames, 
some stacks are very deep (maximum 255 frames). 

Frames in a stack may contain methods from multiple binaries when one binary calls methods from another. Figure~\ref{fig:crashStackProperties} (b) shows the distribution of distinct binaries appearing in a stack (average $\approx 4$). Multiple binaries in a stack indicate that crashes can happen in binaries outside of the entry binary of an application.

\Paragraph{Stack depth and software types.} Stack depth varies a lot across software components and their types. For example, device drivers usually have smaller stack depth (average  $\approx 7$) than applications (average $\approx 11$) and systems (average $\approx 18$). This is most likely because, compared to applications/systems, drivers are less complex in terms of the number of dependencies, and hence tend to make fewer nested method calls.  Figure~\ref{fig:crashStackProperties} (c) shows the average stack depths of 25 most frequent software components. The average depths differs significantly (from 1 to 35) across these software components.  

\myInsight{Finding \#2}
{Stack depths significantly differ across software components and their types.}

\subsection{Problem classes and blamed frames}
\begin{figure*}[!h]
\centering
\subcaptionbox{Top-15 problem classes in crashes\label{fig:probClass}}{\includegraphics[height=0.23\textwidth, width=0.275\textwidth]{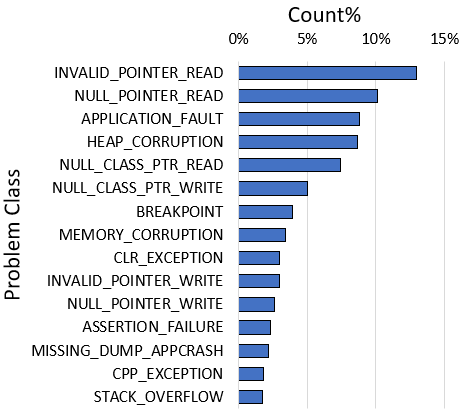}}%
\hfill % <-- Seperation
\subcaptionbox{Normalized crash location\label{fig:crashlocation}}{\includegraphics[height=0.2\textwidth, width=0.275\textwidth]{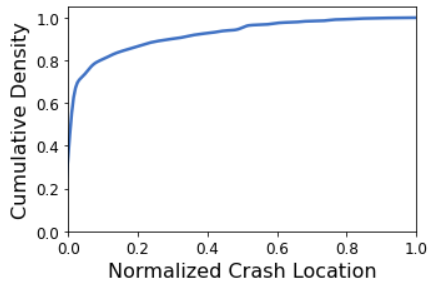}}%
\hfill % <-- Seperation
\subcaptionbox{Distribution of blame-ratio \label{fig:blameratio}}{\includegraphics[height=0.2\textwidth,width=0.275\textwidth]{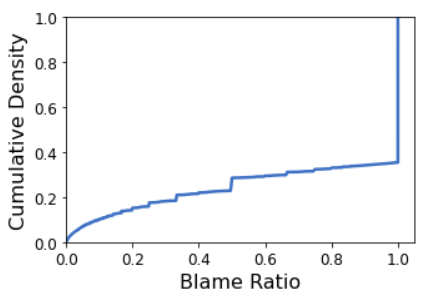}}%
\vspace{-0.5em}
\caption{Properties of problem classes and blame frames}
\vspace{-2mm}
\label{fig:blameProperties}
\end{figure*}

\Paragraph{Problem classes.}
Figure~\ref{fig:probClass} shows top 15 classes of problems that cause the crashes.  Most of the key problem 
classes are memory related, for instance, when the application is trying to read a pointer that points to invalid memory (INVALID\_POINTER\_READ) or is trying to read a pointer that is null (NULL\_POINTER\_READ). These memory-related classes account for $61\%$ of all crashes. Another major problem category is APPLICATION\_FAULT, which is the default class when no more specific class could be identified.
\\
\myInsight{Finding \#3}
{Most ($61\%$) of the crashes are caused by memory-related errors.}

%\suman{What else can we say here about the data?  Or can we put this to some other section?}

\Paragraph{Blamed frame location.} Figure~\ref{fig:crashlocation} shows 
the distribution of {\em normalized location} 
of a blamed frame in its crash stack. Locations are normalized so that the top frame and the bottom frame in a stack have locations 0 and 1 respectively. 
As shown, frames at the top of stacks are more likely to be blamed. 
More specifically, the topmost frame is indeed the blamed frame in $67\%$ crash stacks. In the remaining $33\%$ cases, however, blamed frame is not the top frame. An example is shown in Figure~\ref{fig:example-stack} (a) where the third frame is the blamed frame, and top two frames correspond to harmless logging methods. Also, in $5\%$ cases, blamed frame is at the bottom half of the stack.

\smallskip
\myInsight{Finding \#4}
{Blamed frames are more likely to be located at the top of the stack. In $33\%$ cases, however, blamed frame is below the top frame.}

\Paragraph{Context dependence.} Can a method, once identified as the blamed 
method in a crash stack, always be blamed in other crash 
stacks? If yes, one could easily identify the blamed method in a new crash stack by matching its methods with a list of known blamed methods ({\em blame-list}). Does this simple blame-list-based approach work?

To answer this, we compute {\em blame ratio} of methods. The blame ratio of a method, is the ratio of the number of crash stacks where the method is the blamed method, to the number of crash stacks containing the method. Figure~\ref{fig:blameratio} shows the distribution of blame ratio in our dataset. While a large fraction of methods have a blame ratio of 1, a significant fraction of methods have blame ratios less than 1. A ratio less than 1 means that a blamed method in one stack may not always be blamed in other crash stacks, and hence a blame-list-based approach would not work for these methods.

Why do some methods have blame ratios less than 1? A closer examination reveals that whether a method is indeed a blamed method, often depends on its {\em context} -- methods that appear in frames above and below it. Let's consider the example in Figure~\ref{fig:example-stack} again. Here, the method \texttt{ReleaseDrawTexture} is identified as the blamed method in the first example where it appears below two relatively-harmless logging methods. On the other hand, the same method is not blamed in the second example where more crash-prone driver methods appear above it.

The blamed frame's dependence on context can be explained by the fact that frames in a stack are not independent; rather, they are ordered based on the caller-callee relationship of methods in the frames. The data suggests that whether a method is blamed or not depends on the call chain through which it is invoked.

\smallskip
\myInsight{Finding \#5}
{Whether a method is blamed or not often depends on the context it appears in, i.e., methods that appear above and below it in the stack.}
\vspace{-2mm}

%///////////////////////////////////////
%show data to show what features of stack traces are important.
%\begin{itemize}
%	\item Blame frames tend to be at the top of the stack
%	\item Blame frames tend to be in application code
%	\item Blame frames tend to have good tf-idf scores
%\end{itemize}

% \manish{Adding a bunch of stats about the dataset:}
%\begin{itemize}
%	\item Total crashes = 362249
%	\item Time range = 1 week
%	\item Sampling = 10\% uniform sampling
%	\item unique applications = 8783
%	\item unique binaries = 16259
%	\item unique namespaces = 38153
%	\item unique methods = 85477
%	\item min frames per stack = 1
%	\item max frames per stack = 255
%	\item Mean frames per stack = 16
%	\item Median frames per stack = 9
%	\item unique blamed binaries = 13244
%	\item unique blamed methods = 17673
%\end{itemize}

%\manish{Environment for Insights:}
%\myInsight{Finding \#1}
%{Insights, details, or statement}

% \subsubsection{[\textbf{Crash Stack}] How diverse are crash stacks?}\hfill

% \subsubsection{[\textbf{Crash Binary}] Which binaries and their types predominantly cause crashes?}\hfill

% \subsubsection{[\textbf{Problem Class}] How diverse are crash problems?}\hfill

% \subsubsection{[\textbf{Stack Depth}] How deep are crash stacks?}\hfill

% \subsubsection{[\textbf{Crash Location}] How deep can the crash location be?}\hfill
\section{Our Approach}\label{sec:model}
In this section, we use the insights from Section~\ref{sec:empirical-analysis} to design a data-driven solution to the large-scale crash localization problem. 

Our goal is to utilize large-scale historical crash data, consisting of crash stacks and their labelled blamed frames, to learn models that given a new crash stack can identify its blamed frame (and method). Our first attempt was to learn a linear binary classifier (e.g., Logistic Regression) that, given various features of an individual frame, predicts the likelihood of it being the blamed frame. As we show in 
Section \ref{sec:rq1-eval}, this simple model, however, did not work well for many problem classes. This is explained by one of our findings in Section~\ref{sec:empirical-analysis}: the context in which a method appears in a crash stack (e.g., methods that appear before and after it) plays an important role for it to be a crash method or not. We therefore aim to build models and features that can effectively capture such context. We achieve this with a novel solution that formulates the crash-localization task as a sequence labeling task, as described next. 

\subsection{Crash localization as sequence labeling}
Sequence labeling, well explored in Natural Language Processing (NLP) \cite{nguyen2007comparisons}, involves assigning a categorical label to each member of a sequence of observed values. For example, Named-Entity Recognition\cite{shetty2021neural, ratnaparkhi1996maximum, li2020survey}, a sequence labeling task, can locate and label entities in a sentence as predefined categories. It treats a sentence as a sequence of words and considers context, i.e., surrounding words, of each word to identify its category. For example, in ~\ref{tab:seq-labeling}, the named entity \textit{John} is identified as a \textit{Person} and \textit{Seattle} a \textit{Location}. 

\begin{figure}[!t]
    \begin{tabular}[t]{ll}
    \toprule
         \textbf{Natural Language Processing} & \textbf{Crash Dump Analysis}\\
    \midrule
        \textbf{Sentence}: A sequence of \textbf{words} &
        \textbf{Stack}: A sequence of \textbf{frames}\vspace{0.4em} \\
        
        \begin{tikzpicture}[
            node distance = 0pt,
            square/.style = {draw=blue!90, fill=blue!5, dashed, 
                             minimum height=1.6em, minimum width=3em,
                             outer sep=0pt},    
                ]
            %Nodes
            \node[square]   (nlp0) {John};
            \node[square, right=of nlp0] (nlp1) {lives};
            \node[square, right=of nlp1] (nlp2) {in};
            \node[square, right=of nlp2] (nlp3) {Seattle};
        \end{tikzpicture} & 
        
        \begin{tikzpicture}[
            node distance = 0pt,
            square/.style = {draw=blue!90, fill=blue!5, dashed, 
                             minimum height=1.6em, minimum width=3em,
                             outer sep=0pt},    
                ]
            %Nodes
            \node[square]   (cd0) {$f_{0}$};
            \node[square, right=of cd0] (cd1) {$f_{1}$};
            \node[square, right=of cd1] (cd2) {$f_{2}$};
            \node[square, right=of cd2] (cd3) {$f_{3}$};
        \end{tikzpicture}\vspace{0.3em} \\
    
        \textbf{Sequence Labeling} & \textbf{Crash Localization} \vspace{0.3em}\\
        
        \begin{tikzpicture}[
            node distance = 0pt,
            square/.style = {draw=blue!90, fill=blue!5, dashed, 
                             minimum height=1.6em, minimum width=3em,
                             outer sep=0pt},
            greensquare/.style = {draw=blue!90, fill=green!40, dashed, 
                 minimum height=1.6em, minimum width=3em,
                 outer sep=0pt},
            pinksquare/.style = {draw=blue!90, fill=pink!60, dashed, 
                 minimum height=1.6em, minimum width=3em,
                 outer sep=0pt}, 
                ]
            %Nodes
            \node[greensquare]   (nlp0) {John};
            \node[below=of nlp0] {\textbf{PER}};
            
            \node[square, right=of nlp0] (nlp1) {lives};
            \node[below=of nlp1] {O};
            
            \node[square, right=of nlp1] (nlp2) {in};
            \node[below=of nlp2] {O};
            
            \node[pinksquare, right=of nlp2] (nlp3) {Seattle};
            \node[below=of nlp3] {\textbf{LOC}};
        \end{tikzpicture} & 
        
        \begin{tikzpicture}[
            node distance = 0pt,
            square/.style = {draw=blue!90, fill=blue!5, dashed, 
                             minimum height=1.6em, minimum width=3em,
                             outer sep=0pt},
            redsquare/.style = {draw=blue!90, fill=red!40, dashed, 
                 minimum height=1.6em, minimum width=3em,
                 outer sep=0pt},
                ]
            %Nodes
            \node[square]   (cd0) {$f_{0}$};
            \node[below=of cd0] {!BF};
            
            \node[square, right=of cd0] (cd1) {$f_{1}$};
            \node[below=of cd1] {!BF};
            
            \node[redsquare, right=of cd1] (cd2) {$f_{2}$};
            \node[below=of cd2] {\textbf{BF}};
            
            \node[square, right=of cd2] (cd3) {$f_{3}$};
            \node[below=of cd3] {!BF};
        \end{tikzpicture} \\
    \bottomrule
    \end{tabular}
    \vspace{-2mm}
    \caption{Analogy between NLP and Crash Dump Analysis}
    \vspace{-3mm}
    \label{tab:seq-labeling}
\end{figure}

% \vspace{-1em}

\smallskip
To formulate crash localization as sequence labelling, we consider a crash stack as a sequence of frames, analogous to a sentence and its constituent words. We then perform sequence labelling with a binary category label - \textit{BlameFrame} and  \textit{!BlameFrame}. Thus, the problem is formulated as follows: {\em given a crash stack (i.e., a sequence of frames), label each frame with whether it is a blame frame or not}. For example, in Figure~\ref{tab:seq-labeling}, the third frame ($f_{2}$) is identified as the \textit{BlameFrame} (BF), while the rest are labeled \textit{!BlameFrame} (!BF).
  
For traditional sequence labeling, one can use existing models that have been proposed previously. However, applying them to crash localization requires addressing some unique challenges.
\begin{itemize}
    \item {\em What features to use?} To accurately summarize a crash stack, features need to capture both semantics and domain-specific information.
    So, we make use of Tf-Idf \cite{luhn1957statistical, jones1972statistical} based features, as well as features highlighted by our empirical study, such as frame-depth, that are strongly correlated to crash locations.
    
    \item {\em How to blame exactly one frame?} In traditional sequence labeling, it is possible for multiple tokens to be labeled as the same category; e.g., sentence with multiple places. In a crash stack, however, there is only one blame frame, and hence, we need appropriate models that satisfy such constraints. 
\end{itemize}

In the rest of the section, we describe the design and architecture of our \deepanalyze{} model that addresses these challenges.

\vspace{-1mm}
\begin{table*}
    \caption{Featured to represent stack frames for crash localization}
    \vspace{-2mm}
    \label{tab:features}
    \begin{tabular}[t]{lll}
    \toprule
         \textbf{Feature Group} & \textbf{Feature Name} & \textbf{Description} \\
    \midrule
        Semantics & \texttt{namespace} & $n$ dimensional Tf-Idf vector of the namespace\\
            & \texttt{method} & $n$ dimensional Tf-Idf vector of the method\\
    \midrule
        Types of Code & \texttt{is\_appname\_in\_frame} & Does the frame contain the application's name?\\
            & \texttt{is\_first\_app\_frame} & Is it the $1^{st}$ frame with the application's name?\\
            & \texttt{is\_kernel\_code} & Does the frame contain kernel code?\\
            & \texttt{is\_ntdll\_code} & Does the frame contain \texttt{ntdll} code? \\
            & \texttt{is\_exception\_in\_frame} & Does the frame contain an exception?\\
    \midrule
        Other & \texttt{norm\_frame\_position} & Normalized position of the frame\\
            & \texttt{is\_method\_unknown} & Is the method \texttt{unknown}?\\
            & \texttt{is\_method\_empty} & Is the method empty?\\
            & \texttt{is\_binary\_unknown} & Is the binary \texttt{unknown}?\\
            & \texttt{is\_empty\_frame} & Is the entire frame empty?\\
    \bottomrule
    \end{tabular}
    \vspace{-2mm}
\end{table*}

\subsection{Model Features}
\label{sec:model-features}

Guided by domain expertise and our empirical study, we engineered features shown in Table \ref{tab:features}, for data-driven models. These features transform individual frames in the crash stack into real-valued vectors that can be used by our models. The features are generic, they apply to crashes across applications, and can be grouped into the 3 broad types briefly described below:

\Paragraph{Semantics:} These features represent the important contents of a frame such as the namespace and method name. Here, we observe that tools such as \banganalyze \cite{glerum2009debugging} utilize a large list of allow-lists and heuristics to localize crashes deeper in the stack. Such approaches do not consider the global \textit{semantics} and \textit{relevance} of the function in a frame, i.e., how a function contributes to the crash. To include these semantics, we utilize a simple Term Frequency - Inverse Document Frequency (Tf-Idf) vectorization method \cite{luhn1957statistical, jones1972statistical}. With this approach we automatically extract a weighted list of important tokens from namespaces and methods in frames.
    
\Paragraph{Types of Code:} Code from applications (1$st$ and $3rd$ party) are more likely to have bugs and cause crashes, when compared to to kernel code and core OS user-mode code \cite{glerum2009debugging}. To incorporate such information, we use features that check the presence of the application's name in the frame (i.e. the binary name). We also extract features that represent kernel code, core OS modules, and exceptions. These features can help models de-prioritize frames that are less likely to contain the root cause for crashes.
    
\Paragraph{Other Information:} As shown in Section \ref{sec:empirical-analysis}, frames at the top of stacks are more likely to be blamed. We thus utilize the normalized frame position to model how deep in the stack the crash location can be. Also, at times frames can be incomplete or have missing symbols in scenarios such as some $3rd$ party software, and Linux OS components or standard libraries. To de-prioritize such frames, we use multiple boolean features that check for unknown and missing information in the frame.

By transforming frames using the features described in Table \ref{tab:features}, a stack can now be visualized as a sequence of featurized frames.

\subsection{\deepanalyze{} Model}
\label{sec:deepanalyze-model}

In the following subsections we describe the important components of our multi-task model, as shown in Figure \ref{fig:deepanalyze-model}, in detail.

\subsubsection{\textbf{Model Overview}}
%Machine learning models require real valued vectors representing features of the input to learn patterns. As described in Section \ref{sec:model-features}, we utilize features that represent both semantics and metadata of frames to encode stacks into a sequence of input vectors. With featurized frames available, we now move on to designing the \deepanalyze deep learning model shown in Figure \ref{fig:deepanalyze-model}. 
As our empirical analysis in Section~\ref{sec:empirical-analysis} shows, whether a stack frame is blamed or not often depends on its context -- frames above and below it.
%To accurately model crashes, we first revisit the properties of a stack and its frames. Stacks contain multiple ordered frames that often depict the call graph of methods. This means that a frame is associated with information from both, the frame below it (caller) and above it (callee). 
Hence, while modeling, we need to consider context flowing in both directions in the stack -- top-down and bottom-up. For this, we utilize a Bi-directional LSTM (Bi-LSTM), that interprets the stack, both forwards and in reverse.

While, the BiLSTM can model sequential context flow, dependencies between frames can be widely distributed in the stack. Also, as shown in Section \ref{sec:empirical-analysis}, we observe that stacks can be very long, and BiLSTMs can sometimes fail to handle long-range dependencies \cite{vaswani2017attention}. To overcome these challenges, we implement an Attention mechanism. It favours the model to attend to sections of the stack more likely to have the crash location.

With a Bi-LSTM and Attention layer, \deepanalyze{} encodes frames and stacks into robust neural representations that can be used to localize crashes. Here, we see that unlike sequence labeling for natural language, there is a constraint where we can only label a single frame in the stack as the blame frame. To learn such structural constraints, we use a Conditional Random Fields (CRF) layer. It is a discriminative classifier that models decision boundaries between labels in a sequence.

Lastly, context for crash localization can also be external information that complements the stack. Specifically, symptoms (problem classes) associated with a crash, such as \texttt{Invalid\_Pointers}, \texttt{Zero\_Division}, and \texttt{Heap\_Corruption}. For instance, if the problem was a \texttt{Stack\_Overflow} caused by tail recursion, then the stack would contain a repeating sequence of frames. In this case, the crash can be quickly localized by attending to the repeating pattern. Based on these insights, we utilize multi-task learning to perform \textit{problem class prediction} alongside the primary task - \textit{blame frame prediction}.

%Figure~\ref{fig:deepanalyze-model} shows the architecture of \deepanalyze{} model.

\subsubsection{\textbf{Bi-directional LSTM}}
Long Short-term Memory (LSTM) networks \cite{hochreiter1997long} are a type of Recurrent Neural Networks (RNNs) that have been widely used to process sequential data in a variety of tasks such as language modelling \cite{mikolov2010recurrent, sundermeyer2012lstm}, speech processing \cite{sak2014long}, and code comment generation \cite{hu2018deep}. 
% The LSTM architecture allows it to capture long range dependencies using several gates. 
It takes a sequence of inputs $(x_1, x_2, \dots, x_n)$ as and return a sequence of vectors $(h_1, h_2, \dots, h_n)$ that encodes information at every time step (i.e., frame level here). In our scenario, a frame $f$, receives context from other frames that occur on either sides. We achieve this representation using a second LSTM layer interpreting the sequence in reverse, i.e., a Bi-Directional LSTM (Bi-LSTM) \cite{graves2005framewise}. Finally, each frame is represented by concatenating its left and right context, $h_{f}=[\overrightarrow{h_{f}};\overleftarrow{h_{f}}]$.

\begin{figure}[t]
\includegraphics[scale=0.65]{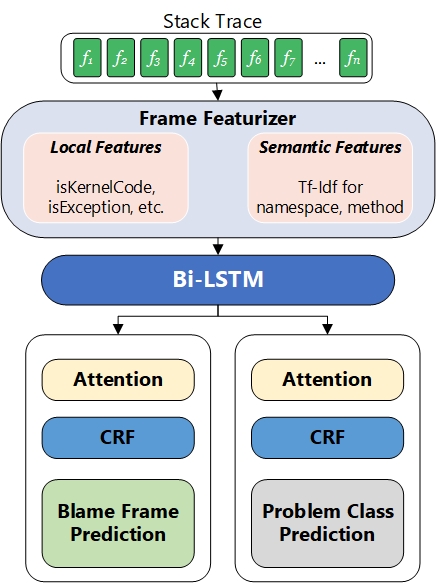}
\vspace{-2mm}
\caption{\textbf{\deepanalyze{} Multi-Task Model Architecture}}
\label{fig:deepanalyze-model}
\vspace{-4mm}
\end{figure}

\subsubsection{\textbf{Attention Mechanism}}

Attention mechanism \cite{bahdanau2014neural, luong2015effective} has become a key component of state-of-the-art solutions to quantify distributed dependencies in sequences. It has been used for tasks like machine translation \cite{bahdanau2014neural}, sentiment classification \cite{chen2017recurrent}, parsing \cite{li2016discourse}, and even image classification \cite{wang2017residual}. 
Here, we implement attention mechanism at the frame level with a learnable parameter $W_a$, as described in Equations \ref{eq:attn-score}-\ref{eq:attn-context-vector}. It takes as input the hidden states $h = [h_1, h_2, \dots,h_{T}]$ from the Bi-LSTM, and generates a weighted context vector $h^*$ of the stack. This weighting mechanism urges the model to focus on sections of the stack that are more likely to have the crash location.

\vspace{-1em}
\begin{equation}
    \label{eq:attn-score}
    scores = W_a^Th
\end{equation}
\begin{equation}
    \alpha = softmax(scores)      
\end{equation}
\begin{equation}
    \label{eq:attn-context-vector}
    h^* = \tanh(h\alpha^T)    
\end{equation}

\subsubsection{\textbf{Conditional Random Fields}}

The above discussed layers encode stack information into neural representations. Next, we move on to labeling the crash location. Here, we could simply predict labels independently for each frame. But we observe that this disregards some structural constraints in our problem. Specifically, unlike traditional sequence labeling, we can only label one frame as the crash location. To enforce such restrictions, we model the frame level labeling task jointly using linear chain conditional random fields (CRF) \cite{lafferty2001conditional}. Given an input sequence $X$, the CRF layer computes the probability of observing an output label sequence $y$, i.e., $p(y|X)$:

\begin{equation}
    s(X,y) = \sum_{i=0}^{n} A_{y_i,y_{i+1}} + \sum_{i=0}^{n} P_{i,y_{i}} 
\end{equation}

\begin{equation}
    p(y|X) = 
    \dfrac{e^{s(X,y)}}{\sum_{y' \in Y} e^{s(X,y')}  }
\end{equation}

Here, $P$ is a probability matrix of shape $n \times k$ from the attention layer, where $k$ is the number of distinct tags and $n$ is the sequence length. $A$ represents the matrix of scores for transitions between output labels. Finally, to extract labels, the layer predicts the output sequence with the highest probability - $y^* = \argmax_{y' \in Y} p(y'|X)$. With this approach the model learns to include structural validity in predicted output sequences.

\subsubsection{\textbf{Multi-Task Learning}}

Multi-Task Learning (MTL) is an approach to improve generalization in models using the inductive bias in jointly learning related tasks \cite{caruana1997multitask}. In the context of classification and sequence labeling, MTL improves performance of individual tasks by learning multiple tasks simultaneously \cite{shetty2021neural}.

In our scenario, the primary task for \deepanalyze{} is crash localization. As stated before, we observe that localizing crashes not only depends on the frames, but also on the class of problems that might have caused the crash. Consequently, we choose problem class prediction as a secondary task for our multi-task model. Particularly, as shown in Figure \ref{fig:deepanalyze-model}, we utilize a multi-head architecture to share low level features (BiLSTM layer). Then the architecture splits into 2 task specific branches - one for blame frame prediction and the other for problem class prediction. The model is trained jointly to minimize a combined loss function, but task specific losses are computed to update individual branch weights.

\subsection{Cross-Application Crash Localization}
\label{sec:cross-application}
As stated in Section \ref{sec:background}, hardcoding heuristics into code creates challenges with scale and generalizability for unknown future scenarios. Software constantly evolves as new applications, APIs, and programming languages are introduced and become popular. Handling crashes in such new cases usually requires a lot of time and deep domain knowledge to write custom rules and plugins for existing solutions. Here, learning a model instead, can help address the scalability and generalizability challenges with ever growing software.

But, even with supervised machine learning, for a new application, it is nontrivial to develop accurate crash localization models, as there would be minimal labeled training data. However, in crashes, we believe that there are many patterns to be learnt that are common across applications; especially the large portion of frames that represent the underlying system (see Figure~\ref{fig:crashingbinaries}(b)). This implies that models trained on crashes from a global set of applications (source) can be used to localize crashes for a new and disjoint set (target) — \textit{Cross-Application Crash Localization}. 

In this work, we use this to overcome the above mentioned challenges with a \textbf{Transfer Learning} and \textbf{Fine-tuning} approach. Formally, following Pan and Yang \cite{pan2009survey}, transfer learning involves the concepts of a domain and a task. 
A domain $\mathcal{D}$ consists of 2 parts; a feature space $\mathcal{X}$ and a marginal probability distribution $P(X)$ where $X = x_1, \dots, x_n \in \mathcal{X}$. 
Given a specific domain, $\mathcal{D} = \{\mathcal{X} , P(X)\}$, a task $\mathcal{T}$ has 2 components; a label space $\mathcal{Y}$ and an objective $f(\cdot)$ (i.e., $T = \{\mathcal{Y}, f(\cdot)\}$), that can be learned from training data. Given this, transfer learning is defined as:

\Paragraph{\textit{Transfer Learning}:} Given a source domain $\mathcal{D}_S$ and learning task $\mathcal{T}_S$, target domain $\mathcal{D}_T$ and learning task $\mathcal{T}_T$, transfer learning aims to improve the learning of the target predictive function $f_T(\cdot)$ in $\mathcal{D}_T$ using the knowledge in $\mathcal{D}_S$ and $\mathcal{T}_S$, where $\mathcal{D}_S \neq \mathcal{D}_T$ , or $\mathcal{T}_S \neq \mathcal{T}_T$.
\vspace{0.5em}

In our scenario, we observe $\mathcal{D}_S \neq \mathcal{D}_T$, where $\mathcal{D}_S$ is the global set of application crashes and the target $\mathcal{D}_T$ is a new/unseen application's crashes. Specifically, we see that the feature space ($\mathcal{X}$) of the source and target are same, while the marginal probability distributions $P(X)$ are different. This case is generally known as ``Domain Adaptation'' \cite{daume2006domain}. With that in mind, for cross-application crash localization, we first pre-train a \deepanalyze{} model on a large dataset of crashes spanning multiple applications (Global model). This model learns general and common information about crashes. Then, for a new application scenario, we use transfer learning to adapt the weights of this global model to the application of interest with minimal labeled data. In Section \ref{sec:evaluation}, we test our hypothesis and extensively evaluate this approach.
\section{Evaluation}
\label{sec:evaluation}

\Paragraph{Implementation.}
We have implemented \deepanalyze{} and all other machine learning models discussed in this work in Python 3.7.7, with Keras-2.2.4 and the tensorflow-1.15.0 backend. The semantic vectorizers are implemented using the standard tf-idf vectorizer in scikit-learn. For our Bi-LSTM CRF based models, the length of the sequence is limited to a maximum of 255, as collected by WER. Also, the hidden layer size is set to 200 cells along with a dropout of 25\% to prevent overfitting by ignoring randomly selected neurons during training. Further, we use an early stopping mechanism to stop training when model performance on a validation dataset starts to degrade. Lastly, our models are trained on an Ubuntu 16.04 LTS machine, with 24-core Intel Xeon E5-2690 v3 CPU (2.60GHz), 112 GB memory and 64-bit operating system. The machine also has a Nvidia Tesla P100 GPU with 16 GB RAM.

We next consider two settings and evaluate the {\em accuracy} of \deepanalyze: the fraction of test crash stacks for which \deepanalyze{} correctly identifies the blame frame.

\begin{figure*}[!ht]
	\captionsetup[subfigure]{labelformat=empty}
	\subfloat[]{
		{
		    \includegraphics[height=0.25\textwidth, width=0.4\textwidth]{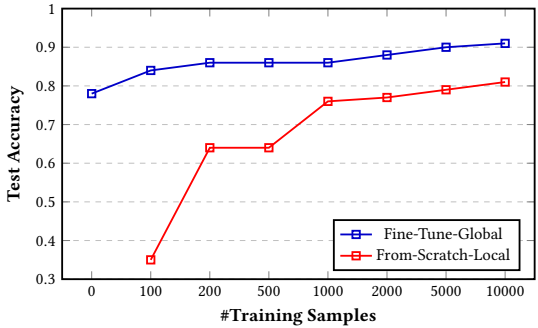}	
	    }}
	\hspace{5em}
	\subfloat[]{
		{
			\includegraphics[height=0.25\textwidth, width=0.4\textwidth]{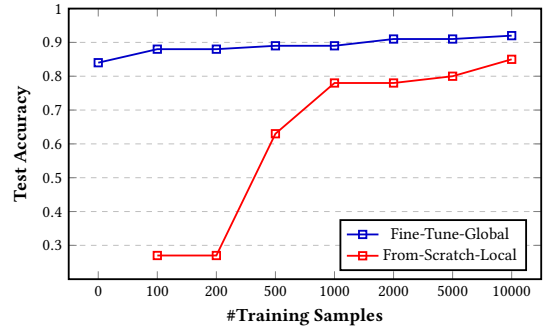}
	    }}
    \vspace{-6mm}
	\caption{Fine-tuning vs Training from scratch for Edge (left) and Excel (right)}
	\vspace{-2mm}
	\label{fig:finetuning}
\end{figure*}

\subsection{Application-Specific Evaluation}
\label{sec:rq1-eval}

We first consider an {\em application-specific} setting where the training and test data come from the same applications. This makes sense when the target application has sufficient labelled training data.

\Paragraph{Setup.} We here use crash stacks from 4 popular client applications from Microsoft - \textit{Edge}, \textit{Excel}, \textit{Word}, and \textit{Outlook}. To evaluate in a realistic setup, we train and test our models at different points in time. We begin by collecting a sample of $\approx1.2$ million user mode crashes of these applications over a window of $2$ weeks. For each application, we utilize data from the first $11$ days for training and next $3$ days for testing (11:3 $\approx$ 80\%:20\%). %Consequently, during evaluation, we ensure \deepanalyze{} only utilizes past data to generalize and localize future unseen crashes. 
Also, we use a combined hash of the stack, blamed frame, and other metadata to de-duplicate our datasets, avoiding multiple evaluations of the same problem. Lastly, we establish ground truth using \banganalyze{} as used by WER.

We compare our multi-task model (described in Section \ref{sec:deepanalyze-model}) against multiple heuristics and machine learning baselines. In Table \ref{tab:eval-app-specific}, we report the accuracy (ratio of $\#$correctly localized crashes to $\#$total crashes) and evaluate models on 4 different applications.

\begin{table}[t!]
    \vspace{-2mm}
    \caption{Evaluation of App-Specific Model Accuracy}
    \vspace{-2mm}
    \label{tab:eval-app-specific}
    \begin{tabular}[t]{lccccl}
    \toprule
    
         \textbf{Model} 
         & \multicolumn{4}{c}{\textbf{Application}} & \textbf{Avg}\\\cmidrule(lr){2-5}
         
         & \textbf{Edge} & \textbf{Excel} & \textbf{Word} & \textbf{Outlook} & \\
         
    \midrule
    
        \texttt{TopFrame} & 0.64 & 0.77 & 0.70 & 0.62 & 0.68 \\
        \texttt{SecondFrame} & 0.24 & 0.07 & 0.10 & 0.13 & 0.13 \\
        \texttt{MostFreqTopFrame} & 0.31 & 0.42 & 0.39 & 0.39 & 0.38 \\
                         
        Logistic Regression & 0.86 & 0.81	& 0.75 & 0.69 & 0.77 \\
        
        BiLSTM-CRF-Attn & 0.91 & 0.90 & 0.80	& 0.81 & 0.85 \\
        \textbf{DeepAnalyze} & \textbf{0.93} & \textbf{0.94} & \textbf{0.85}	& \textbf{0.88} & \textbf{0.90} \\
            
    \bottomrule
    \end{tabular}
    \vspace{-3mm}
\end{table}

\begin{table}[t!]
    \caption{Significance of Improvements}
    \vspace{-2mm}
    \label{tab:improvements}
    \begin{tabular}[t]{p{2cm}ccccl}
    \toprule
   
         \multirow{2}{2cm}{\textbf{Improvement Area}} 
         & \multicolumn{4}{c}{\textbf{Application}} & \textbf{Avg}\\\cmidrule(lr){2-5}
         
         & \textbf{Edge} & \textbf{Excel} & \textbf{Word} & \textbf{Outlook} & \\
         
    \midrule
    
        \textit{Semantics}
            & 29\% & 5\% & 6\% & 10\% & 13\% \\
        \textit{Context} & 8\% & 15\% & 13\% & 24\% & 15\% \\
        \textit{Multi-Task} & 3\% & 4\% & 6\% & 9\% & 5\% \\
        
    \bottomrule
    \end{tabular}
    \vspace{-3mm}
\end{table}

\Paragraph{Heuristic Baselines.} First, we have a \texttt{TopFrame} baseline that always picks the $1^{st}$ frame in the stack. This is based on the insight that a large proportion of crash locations are at the top of the stack. But, we also observed that in some cases the top frame is an exception raised by the method below it. We represent this using our \texttt{SecondFrame} baseline that always picks the $2^{nd}$ frame. Next, with \texttt{MostFreqTopFrame}, we introduce the use of frequent patterns. This baseline picks the frame that was most frequently blamed in the past. In case of ties or unseen frames, it favours frames higher in the stack. From Table \ref{tab:eval-app-specific}, we see that these heuristic approaches perform well only on certain crashes and do not generalize well.

\begin{figure}[!ht]
\vspace{0.25mm}
\includegraphics[scale=0.6]{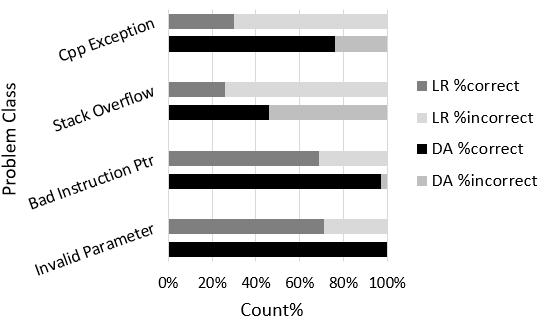}
\vspace{-2mm}
\caption{Logistic Regression (LR) vs \deepanalyze{} (DA)}
\label{fig:logreg-vs-deepanalyze}
\vspace{-4mm}
\end{figure}

\Paragraph{Linear Model.} Next, we have a Logistic Regression baseline. It is linear binary classifier that, given features of an individual frame (described in Section \ref{sec:model-features}), predicts the likelihood of it being the blame frame. Then, we pick the frame in the stack with maximum likelihood. As seen in Table \ref{tab:eval-app-specific}, this simple data-driven model performs better than naive heuristics ($0.77$ avg accuracy).
But, on further analysis, we found that this model performs poorly for specific problem classes. For instance, in Figure \ref{fig:logreg-vs-deepanalyze}, we see that for Cpp Exception and Stack Overflow, it correctly predicts only ~$30\%$ of cases. Here, we observed that such problems have diverse crashes where a method can be blamed in one crash stack but not in another. That is, the logistic regression approach lacks in capturing \textit{context}.

\Paragraph{Sequence Labeling.} Lastly, we evaluate models that incorporate the missing context using our novel sequence labeling formulation -- BiLSTM-CRF-Attn and \deepanalyze{}. Here, we use BiLSTM-CRF with attention mechanism as a baseline as it is a state-of-the-art model for sequence labeling in NLP \cite{luo2018attention, lample2016neural}. As shown in Table \ref{tab:eval-app-specific}, it achieves an average accuracy of around 0.85. Whereas, our \deepanalyze{} multi-task model (described in Section \ref{sec:deepanalyze-model}) achieves a higher average accuracy of 0.90 and beats all baselines across applications. It also reaches a maximum accuracy of 0.94 for Excel. With \deepanalyze{} combining context information and complementary information using multi-task learning, we are able to outperform strong baselines such as Logistic Regression and BiLSTM-CRF-Attn. Also, though in Figure \ref{fig:logreg-vs-deepanalyze} we mention only some problem classes, we find that \deepanalyze{} is always better than logistic regression.

\Paragraph{Improvements.}
Table  \ref{tab:improvements} shows the significance of important aspects of our approach, namely frame semantics, context dependence, and multi-task learning. We compute significance by calculating the percentage difference in accuracy ($A$) of pairs of models; i.e., $|A_{m1} - A_{m2}|/avg(A_{m1}, A_{m2}) \times 100\%$.
To capture significance of \textit{Semantics}, we compare Logistic Regression, that uses semantic features, with the naive \texttt{TopFrame} baseline. As shown, introducing frame features/semantics, generates considerable improvements across applications (average $13\%$).
Next, we evaluate the value of \textit{Context}. Here, we compare \deepanalyze, a context-aware approach, to Logistic Regression. The $8\%-24\%$ boosts achieved with our approach highlights the importance of context in crash localization. Also, incorporating context provides the largest gains on average ($15\%$). Lastly, \deepanalyze{} leverages both context and complementary information using multi-task learning. Thus, by comparing the multi-task \deepanalyze{} model with BiLSTM-CRF-Attn, we see that multi-task learning also provides notable increases in accuracy (average $5\%$).

\Paragraph{What does \deepanalyze{} learn?} To gain insight into what \deepanalyze{} learns, we use a model weight inspection technique. It is commonly used to interpret black-box models in image processing \cite{karpathy2014large, lee2009convolutional} and medical domains \cite{mehrabi2015temporal}. Here, we extract the weights of the $1^{st}$ layer of \deepanalyze, where there is a direct interaction on the raw inputs, to generate normalized feature importances. Table \ref{tab:feature-importance} summarizes the top-3 positive and negative features, all of which are tf-idf semantic features. As shown, \deepanalyze{} intelligently learns that methods performing memory, file, and thread operations are positive features as they work with pointers and tend to cause crashes. This is supported by our empirical analysis in Section \ref{sec:empirical-analysis} on frequent problem classes. On the other hand, \deepanalyze{} also learns to negatively associate standard libraries (namespace std) and methods that raise errors/exception with crash locations, supporting our observation in Figure~\ref{fig:crashingbinaries}(b) that crashes are relatively less frequent in libraries.  

% _namespace_std = -1.0
% _method_error = -0.759
% _method_exception = -0.59
% _method_memory = 1.0
% _namespace_file = 0.734
% _method_thread = 0.4963
% is_first_app_frame = 0.43

% \vspace{-0.25mm}
\begin{table}[!t]
    \caption{Feature Importances}
    \vspace{-2mm}
    \label{tab:feature-importance}
    \begin{tabular}[t]{ll|ll}
    \toprule
        \textbf{+ve feature} & \textbf{Imp} & \textbf{-ve feature} & \textbf{Imp} \\
    \midrule
        method \texttt{memory} & 1.0 & namespace \texttt{std} & -1.0 \\
        namespace \texttt{file} & 0.73 & method \texttt{error} & -0.76 \\
        method \texttt{thread} & 0.49 & method \texttt{exception} & -0.59\\
    \bottomrule
    \end{tabular}
    \vspace{-3mm}
\end{table}

%%%%%%%%%%%%%%%%%%%%%%%% RQ 2 %%%%%%%%%%%%%%%%%%%%%%%%

\subsection{Cross-Application Evaluation}
\label{sec:rq2-eval}
We now evaluate \deepanalyze{} in a {\em cross-application} setting where we have a recent/new application with minimal labeled crashes. For this, we attempt to evaluate the efficacy of our transfer learning approach for cross-application crash localization (Section \ref{sec:cross-application}).
%To utilize \deepanalyze{} for new and recent applications, in Section \ref{sec:cross-application}, we proposed a transfer learning based approach for cross-application crash localization. 
To summarize, we hypothesis that models learnt from crashes of a set of applications can localize crashes in new/unseen applications. Here, we evaluate our hypothesis on 2 target applications -- \textit{Edge} \& \textit{Excel}. 

\Paragraph{Setup.}
We start with the dataset used in Section \ref{sec:empirical-analysis}, consisting of  $\approx 362K$ crash stacks from many software components, sampled over a period of 1 week. To simulate a cross-application setting, we choose a target application and remove all its associated crashes from our dataset. Then, we train a \deepanalyze{} model (Global Model) on the resultant dataset and test on crashes of the target (domain transfer). Further, we fine-tune our global model to the target application with minimal labeled crashes. To evaluate, we compare this transfer learning approach against an application-specific \deepanalyze{} model (Local Model). 
% Note that, again, we only evaluate on crashes from the week following that of the training data. 

Figure \ref{fig:finetuning} show the results of our experiments for 2 target applications - Edge (left) and Excel (right). The \texttt{Fine-Tune-Global} (blue) line indicates the accuracy of our global model on fine-tuning (i.e., our transfer learning approach). The \texttt{From-Scratch-Local} (red) line indicates the accuracy of an application-specific model trained from scratch. The X-axis has the number of samples used for fine-tuning the global and training the local model, respectively. 

\Paragraph{Accuracy.}
As seen, for both Edge and Excel, our transfer learning approach significantly outperforms a local model trained from scratch, at all amounts of training data. This is mainly because a global model receives a significant head start by learning signals and patterns in crashes that are application agnostic, and hence, transferable. Specifically, we observe that our global model achieves high accuracies ($0.78$ for Edge; $0.84$ for Excel), without observing a single application-specific crash (@ 0 training samples). This shows that \deepanalyze{} can indeed learn from a global crash dataset to effectively localize crashes in new/unseen scenarios. Also, we observe that our models gradually improve with minimal labeled data. For instance, using transfer learning, we achieve $\approx 0.90$ accuracy with as few as 1000-2000 samples, for both Edge and Excel. This encourages that our transfer learning approach can be used in a real-word setting, where for a new application at first we directly use a global model. But, over time, we would collect app-specific labeled data and improve our models. 

\Paragraph{Cost Savings.}
Furthermore, during our experiments we observed that transfer learning not only reduces training data requirements, but also training time compared to application-specific models. To evaluate, we make use of our Edge and Excel models trained on large datasets in Section \ref{sec:rq1-eval}. Note, these models are comparable to the fine-tuned models because they are tested on the same set of crashes. These models took on average $3$ hours to train and received an average accuracy of $0.93$. On the other hand, from Figure  \ref{fig:finetuning}, comparable global models ($\approx 0.91$ acc with 5000-10000 samples) took on average $10$ minutes to be fine-tuned. That is, we see nearly $18X$ reduction in training time, and thus compute cost, with a transfer learning approach. This is particularly encouraging of the usability of such an approach to quickly develop accurate models for newly deployed scenarios.
\section{Related Work}
%\subsection{Crash Analysis}
\Paragraph{Crash Analysis.}
Debugging and triaging of crashes at scale can be expensive and intractable. Our work is most closely related to prior work on large-scale analysis of real world software crashes. The Windows Error Reporting (WER) \cite{glerum2009debugging} distributed system was built by Microsoft for collecting and analyzing crash traces. With \deepanalyze{}, we leverage a novel machine learning based approach for crash localization using the crash traces collected by WER in the wild. Our approach not only generalizes well for several existing applications but can also extend to new applications with very limited amount of labelled data using transfer learning. Wu et al. \cite{wu2014crashlocator} proposed CrashLocator which uses call stack information in the crash reports along with the static call graph information from the source code to predict the blame probability of each frame in the stack. In DeepAnalyze, since we are doing crash localization at scale in the wild, we don't have access to the source code. So, we only use the call stacks from the crash traces to do the crash localization. Further, we leverage recent advances in the machine learning and NLP domain for crash localization. Crash bucketization is an important part of triaging crash reports. The WER system leverages more than 500 heuristics for bucketing. Dang et al. \cite{dang2012rebucket} proposed ReBucket which uses crash stack similarity to assign them to appropriate buckets. Similarly, TraceSim \cite{vasiliev2020tracesim} leverages TF-IDF and Levenshtein distance on crash reports for measuring similarity of crash traces for better triaging and de-duplication. Our work is complementary to these efforts since more precise crash localization can aid with crash bucketization and triaging.

%\subsection{Multi-task Learning}
\Paragraph{Multi-task Learning.}
Multi-task Learning (MTL) \cite{caruana1997multitask, zhang2021survey} is a well-studied technique in the machine learning community. MTL is used to improve the generalization and performance of ML models on a given task by jointly training on other related tasks. MTL has been utilized in several domains such as NLP \cite{collobert2008unified, liu2016recurrent, subramanian2018learning, mccann2018natural, changpinyo2018multi}, speech \cite{wu2015deep, kim2017joint, chen2015speech, shinohara2016adversarial, anastasopoulos2018tied} and healthcare \cite{bi2008improved, caruana1996using, zhang2012multi, harutyunyan2019multitask, hussein2017risk}. In the NLP domain, Collobert et al. \cite{collobert2008unified} proposed a novel convolutional network architecture which uses MTL to jointly perform several NLP tasks such as POS tagging, named-entity extraction and measuring semantic similarity. 

In the software engineering domain, MTL has been leveraged to a limited extent. Prior work has heavily focused on using MTL for building language models for source code \cite{liu2020self, liu2020multi, wang2021mulcode} and software artifacts like bug reports and discussions \cite{shetty2021neural, shetty2021softner, li2020deep, gong2019joint}. Wang et al. \cite{wang2021mulcode} propose MulCode, a MTL based approach to learn a unified representation of source code by jointly training on three tasks: author attribution, comment classification and duplicate function detection. Their evaluation show the efficacy of MTL by outperforming state of the approaches which addressed these tasks separately. To the best of our knowledge, \deepanalyze{} is the first effort to use MTL in context of debugging. It leverages MTL to jointly perform crash localization and problem class identification from crash stacks. Based on the experiments on several popular applications, MTL significantly boosts the accuracy.
\section{Discussion \& Conclusion}

In this paper, we proposed a novel data-driven solution to address the crash-localization problem at scale. We presented the first large-scale empirical study of $362K$ crashes and their blamed methods reported to WER by many Microsoft applications running in the wild. The analysis provides valuable insights on where and how the crashes happen and what methods to blame for the crashes. These insights enable us to develop \deepanalyze{}, a novel multi-task sequence labeling approach for identifying blamed frames in a stack trace. We evaluate our model with real-world crashes from four popular Microsoft applications and show that \deepanalyze{}, when trained with crashes from one application, can not only accurately localize crashes (with $\approx 90\%$  accuracy)  of the same application, but also bootstrap crash localization for other applications with zero to very little training data. This makes \deepanalyze{} a practical solution to be used in the wild for a large number of applications.

As next step, we are planning to integrate \deepanalyze{} with the WER service along with a feedback loop. Using the feedback provided by developers, we will train \deepanalyze{} in an online learning setting. While in this work we tackle the fundamental problem of crash localization, systems like WER aid in various other tasks. For instance, they also perform crash bucketization and root cause hypothesis testing. Current solutions for these tasks, similar to crash localization, are mostly rule based which does not scale and generalize easily to new scenarios. Lastly, we will also be looking at new problems like inter-crash dump correlation when there are multiple OS running on a single device, such as in gaming consoles. Similarly, cross-platform and cross-OS crash localization in a data efficient manner is also critical. We plan to extend \deepanalyze{} to solve these challenges with the overarching goal of simplifying debugging in the large. 

\iffalse
\section{Data Availability}
Due to confidentiality reasons, we have requested privacy and security approvals for publicly releasing the data and code. If approved, we will strive to release them at the earliest.
\fi

% Define bibliography
\balance
\bibliographystyle{ACM-Reference-Format}
\bibliography{references}

\end{document}